\documentclass[11pt,twoside]{article}
\usepackage{asp2006}
\usepackage{epsf}
\usepackage{psfig}
\usepackage{lscape}
\usepackage{graphicx}

\begin{document}
\title{Variation in p-mode power over solar cycle 23 as seen from BiSON and GOLF observations}   
\author{R. Simoniello$^{1,2}$, D. Salabert$^{2}$, R. A. Garc\'ia$^{3}$}   
\affil{$^{1}$Physikalisch-Meteorologisches Observatorium Davos/World Radiation center, 
Dorfstrasse 33, CH–7260 Davos Dorf, Switzerland}
\affil{$^{2}$Instituto de Astrof\'{\i}sica de Canarias, E-38205 La Laguna, Tenerife, Spain} 
\affil{$^{3}$Laboratoire AIM, CEA/DSM-CNRS-Universit\'e Paris Diderot, CEA, IRFU, SAp, Centre de Saclay, 
F-91191, Gif-sur-Yvette, France}    

\begin{abstract}We analyzed BiSON and GOLF/SoHO data with a new technique, to investigate p-mode power 
variation over solar cycle 23. We found a decrease in the mean velocity power of about 20\% for BiSON 
during the ascending phase, in agreement with previous findings. We also found that GOLF, during the red-wing 
configuration, seems to be working at a different height than the theoretically computed one.
\end{abstract}

\section*{Introduction}
The effect of the magnetic field on solar p-mode power is interesting not only for the understanding of 
the propagation of the waves in the Sun's atmosphere, but also for providing information about the surface 
and the cavity within they propagate. It is known that p-mode power is suppressed in magnetically active 
regions (Woods et al. 1981, Lites et al. 1982, Brown et al. 1992) and the associated mechanism is not fully understood yet. It has been suggested 
that the efficiency of the mode excitation mechanism is reduced in magnetic areas (Goldreich et al. 1977, Goldreich et al. 1988, Cally 1995, Jain et al. 1996),
 but it has also been suggested that the magnetic regions, such as sunspots, are strong absorbers of p-mode power 
(Haber et al. 1999, Jain et al. 2002). Furthermore, there are observational evidences that the acoustic power is enhanced around magnetic 
areas at frequencies above 5mHz (Hindman et al. 1998, Thomas et al. 2000). We aim to study the p-mode power dependence with the 
solar magnetic activity cycle. In this preliminary work, we investigated with a new technique the p-mode power 
variation over solar cycle 23 below 5mHz.

\section{Integrated sun-light measurements}
The {\bf Bi}rmingham {\bf S}olar {\bf O}scillation {\bf N}etwork (BiSON) has been operating since 1976 
and is one of the important players in the field of global helioseismology (Chaplin et al. 1996). BiSON is a network 
composed of six resonant scattering spectrometers that perform integrated sunlight Doppler velocity measurements 
of the K (769nm) Fraunhofer line. The nominal height is at $\approx$260 Km above the photosphere (Jim\'enez-Reyes et al. 2007).

The {\bf G}lobal {\bf O}scillation at {\bf L}ow {\bf F}requency (GOLF) instrument on board the SOHO 
spacecraft is devoted to the search of low-degree modes. It works by measuring the Doppler shifts of the Na line 
at 588.9nm (D1) and 589.6nm (D2) (Gabriel et al. 1995). Due to the malfunctioning of the polarization system, GOLF is working in one-wing 
configuration, that has been changing over 10 years of observations as follows (Garc\'ia et al. 2005): 1)1996-1998 in the 
blue-wing configuration; 2) 1998-2002 in the red-wing configuration; 3) up today in the blue-wing configuration. 
Due to the formation heights of the spectral lines, GOLF observes at $\approx$330 Km and at $\approx$480 Km in the 
blue-wing and red-wing configurations respectively (Jim\'enez-Reyes et al. 2007). 
\begin{figure}
\begin{center}
\includegraphics[scale=0.70]{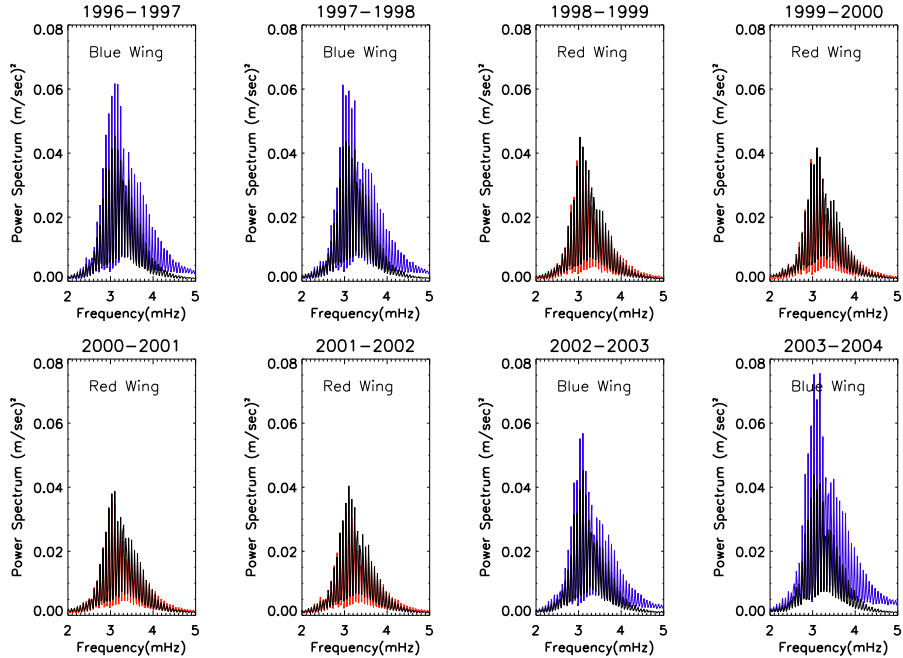}
\caption{Yearly p-mode power spectra from BiSON (black) and GOLF (blue- and red-wing configurations) observations.}
\label{fig:comp}
\end{center}

\end{figure}
\section{Data analysis: Mean Fourier Transform}
We investigated p-mode power variation for the global modes as a whole: this allowed us to split the time series 
in subsets of half-day length and therefore to locate the power in the mode frequency bin. We performed the 
FFT transform for every subset and then we took an yearly average. Figure~\ref{fig:comp} shows the mean Fourier 
spectra obtained with BiSON and GOLF data within the frequency range 2-5mHz. Theoretically, p-mode power increases 
with height in the atmosphere following an exponential behavior (Simoniello et al 2008a, 2008b). Because of the different 
formation heights of the two spectral lines, we expect to observe more power in GOLF than in BiSON observations. 
Instead, Fig.~\ref{fig:comp} shows that from 1998 up to 2000, GOLF signal is lower than for BiSON. Furthermore, 
between the blue- and red-wing configurations, there is a clear evidence of the different contributions from 
the background at low frequency and from the photon-noise level at high frequency (4-5mHz). The photon noise 
is proportional to the counting rate, which is lower in the red-wing configuration, hence less noise. 
In addition, the aging of the instrument affects the counting rate. Since the beginning of the mission, the counting
rate dropped by a factor 10 (Garc\'ia et al. 2005).

\section{P-mode power variation over solar cycle 23}
\begin{figure}
\begin{center}
\includegraphics[scale=0.60]{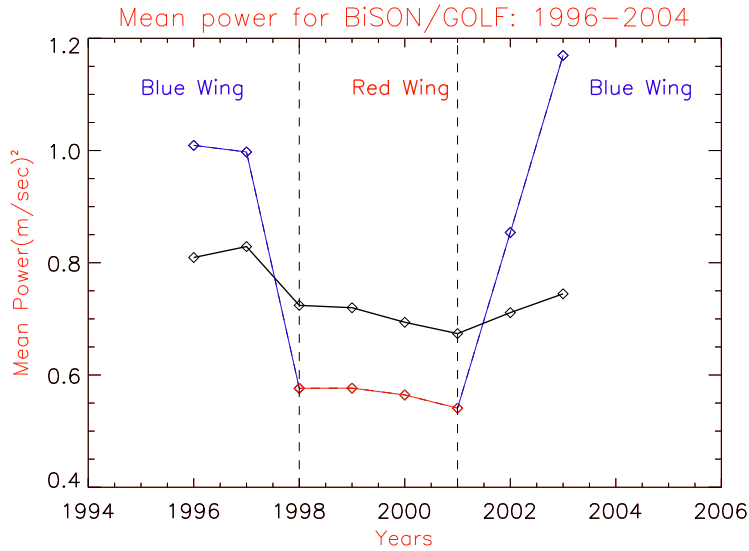}
\caption{P-mode power variation over solar cycle 23 as seen from BiSON (black) and GOLF (blue- and 
red-wing configurations) observations.}
\label{fig:cycle}
\end{center}
\end{figure}

The acoustic p-mode power has been found to decrease of about 20\% over the ascending phase of 
solar cycle 23 (Chaplin et al. 2000, Jim\'enez-Reyes et al. 2003, Salabert et al. 2006). Figure~\ref{fig:cycle} shows a decrease 
in power in both data sets, but with different variations. A decrease of about 20\% is observed in BiSON data, 
which is in agreement with previous findings, while in the case GOLF data, this decrease is about 50\%. 
In 1998, GOLF swapped from blue- to red-wing configuration. This change has shifted higher up the 
observational height (see sec.1.2). Theoretically, GOLF should observe more power during its red-wing configuration
than during its blue-wing configuration and BiSON. This is not what Fig.~\ref{fig:cycle} shows: 
it is a further evidence that GOLF in the red-wing configutation is not observing at the theoretical height.
 
\section{Conclusions}
We applied a new analysis technique that allowed us to investigate and to characterize p-mode power 
variation over solar cycle 23. This investigation confirms the link between p-mode power suppression and 
solar cycle in the region below 5mHz. Furthermore, we showed that GOLF observations, during the red-wing configuration, 
seems to be working at a different height compared to the theoretical one.

\acknowledgements 
This work has been partially funded by the Swiss National Funding 200020-120114, by the Spanish grants PNAyA2004-04462, PNAyA2007-62650 and 
the CNES/GOLF grant at the SAp-CEA/Saclay. 
SOHO is a cooperation between ESA and NASA. 
We thank all the people involved in BiSON (funded by the UK PPARC program).

\end{document}